# Polarization-Orthogonal Nondegenerate Plasmonic Higher-Order Topological States

Yuanzhen Li[1,2,6], Su Xu[3,*], Zijian Zhang[1,2,6], Yumeng Yang[1,2,6], Xinrong Xie[1,2,6], Wenzheng Ye[1,2], Feng Liu[1], Haoran Xue[4,*], Liqiao Jing[1,2,6], Zuojia Wang[1,2,6], Qi-Dai Chen[3], Hong-Bo Sun[3,5], Erping Li[1,2], Hongsheng Chen[1,2,6,7,*], Fei Gao[1,2,6,*]

[1]Interdisciplinary Center for Quantum Information, State Key Laboratory of Extreme Photonics and Instrumentation, ZJU-Hangzhou Global Scientific and Technological Innovation Center, Zhejiang University, Hangzhou 310027, China.

[2]International Joint Innovation Center, The Electromagnetics Academy at Zhejiang University, Zhejiang University, Haining 314400, China

[3]State Key Laboratory of Integrated Optoelectronics, College of Electronic Science and Engineering, Jilin University, 2699 Qianjin Street, Changchun 130012, China.

[4] Division of Physics and Applied Physics. School of Physical and Mathematical Sciences, Nanyang Technological University, Singapore 637371, Singapore.

[5] State Key Laboratory of Precision Measurement Technology and Instruments, Department of Precision Instrument, Tsinghua University, Haidian, Beijing 100084, China.

[6]Key Lab. of Advanced Micro/Nano Electronic Devices & Smart Systems of Zhejiang, Jinhua Institute of Zhejiang University, Zhejiang University, Jinhua 321099, China

[7]Shaoxing Institute of Zhejiang University, Zhejiang University, Shaoxing 312000, China

[*]Authors to whom correspondence should be addressed;

E-mail: *xusu@jlu.edu.cn* (S. Xu), *haoran001@e.ntu.edu.sg* (H. Xue), *hansomchen@zju.edu.cn* (H. Chen), *gaofeizju@zju.edu.cn* (F. Gao).

# Abstract

Photonic topological states, providing light-manipulation approaches in robust manners, have attracted intense attention. Connecting photonic topological states with far-field degrees of freedom (DoFs) has given rise to fruitful phenomena. Recently emerged higher-order topological insulators (HOTIs), hosting boundary states two or more dimensions lower than those of bulk, offer new paradigms to localize/transport light topologically in extended dimensionalities. However, photonic HOTIs have not been related to DoFs of radiation fields yet. Here, we report the observation of polarization-orthogonal second-order topological corner states at different frequencies on a designer-plasmonic Kagome metasurface in the far field. Such phenomenon stands on two mechanisms, *i.e.,* projecting the far-field polarizations to the intrinsic parity DoFs of lattice modes and the parity splitting of the plasmonic corner states in spectra. We theoretically and numerically show that the parity splitting originates from the underlying inter-orbital coupling. Both near-field and far-field experiments verify the polarization-orthogonal nondegenerate second-order topological corner states. These results promise applications in robust optical single photon emitters and multiplexed photonic devices.

Topological physics, originating from originating from solid-state electronic systems, exhibits unique and counterintuitive properties which promise to revolutionize technologies across a variety of fields, from solid-state electronics [1, 2] to photonics [3-5]. Fundamentally different from the quantum systems of equilibrium status, photonic systems with intrinsic nonequilibrium natures can interact with radiation fields [6-15]. Such interactions have brought fruitful consequences ranging from intriguing topological physics to novel device applications by projecting the degrees of freedom (DoFs) of radiation fields to those of topological modes. On the fundamental aspects, linear momentums of radiation fields are projected to in-plane Bloch momentums and have spawned angle-resolved transmission/reflection technologies for investigating underlying topological phenomena, *e.g.*, helicoid Fermi surfaces [6-8], bound states in continuum [9], bulk Fermi arcs [10] or meron spin textures [11]. Practically, polarizations or orbital angular momentums of radiation fields, which are widely utilized in high-capacity wireless communications [16-17] or high-performance quantum sources [18-19], are linked to polarizations or angular momentums of in-plane topological modes and have enabled multiplexed topological radiative devices [13-15].

Recently, a novel class of topological phases, *i.e.*, higher-order topological insulators (HOTIs), have been discovered [20-42]. These emerging topological systems, hosting boundary states two or more dimensions lower than that of HOTIs, offer a new paradigm to localize/transport waves in extended dimensionalities. The exotic properties of HOTIs have been extensively demonstrated in various platforms, including mechanics [23, 31], circuits [22, 24, 41], acoustics [26, 27, 33-35, 37-40], and photonics [21, 28-30, 32, 36]. In further advancements on photonic HOTIs, intense interests are attracted to both multiplexed HOTIs and HOTI radiative devices, such as near-field multiplexed HOTIs realized by synthesizing valley/pseudospins DoFs [43-45], and nonmultiplexed HOTI lasers demonstrated as unusual radiative devices [46, 47]. However, photonic HOTIs have not been related to DoFs of radiation fields yet.

Here, we report the observation of polarization-orthogonal nondegenerate plasmonic second-order topological corner states (SOTCS) in the far field shown as the schematic in Fig. 1. Such exotic phenomenon is based on two crucial factors, *i.e.,* projecting the far-field polarizations to the intrinsic parity DoFs of lattice modes, and parity splitting of the plasmonic SOTCS in spectra. The parity splitting originates from the underlying inter-orbital coupling. The plasmonic HOTI is implemented on a designer-plasmonic Kagome metasurface. By using near-field spectral and imaging techniques, we experimentally observed the parity splitting of SOTCS. Normally illuminating the designer-plasmonic HOTI with linearly polarized plane waves, the far-field experiments demonstrate the polarization-orthogonal SOTCS. Such polarization-orthogonality of nondegenerate SOTCS promises applications in robust optimal single photon emitters [18] and multiplexed photonic devices [19].

We first show the parity splitting of SOTCS in a two-dimensional (2D) Kagome crystal on an airhole-decorated plasmonic media with plasma frequency $\omega_p$ (Figs. 2a-2d) [49]. Each airhole hosts multiple multipolar modes as shown in Fig. 2e. We choose the hexapole modes to work with, since they are commensurate with the $C_3$ symmetry of the Kagome lattice. The hexapole modes in an air hole resonate at frequency $\omega_h$ (termed as zero frequency in the following). The parity (even/odd mode as shown in Figs. 2c-d) is an intrinsic DoF of each site resonance, different from the pseudospin DoFs synthesized with specific crystalline symmetries [43-45]. On a finite-sized triangular lattice with hole distances ($d_1 = 1.21 d_2$) in Fig. 2a, the first-principle results (Figs. 2b-2d) show two sets of parity-dependent topological states. Strikingly, even/odd-parity SOTCS lie above/below the zero frequency, unlike previously demonstrated SOTCS [20-24, 26-47] pinned at zero frequencies.

To understand the parity splitting of SOTCS, we use the tight-binding method to model the plasmonic Kagome lattice. Previously, the correspondence between bulk topology and SOTCS on Kagome lattice has been established by using the Hamiltonian $\hat{H}_{KM}$ based on the nearest-neighbor (NN) couplings of specified *parity-less* orbitals [25-27]. Regarding the *parityful* hexapole at hand, it is straightforward to obtain two copies of parity-dependent Hamiltonian $\hat{H}_{KM}$. The parity-dependent $\hat{H}_{KM}$ results from the parity-dependent NN coupling factors, *i.e.*, $\kappa_{even} = -\kappa_{odd}$, according to fields overlap integral [48, 49]. The hole distances $d_1(d_2)$ are modeled as $\kappa_{h1}$ ($\kappa_{h2}$). Due to the tight confinements of plasmonic modes, the direct long-range coupling (*i.e.,* hole distance of $d_1 + d_2$) between hexapole modes is vanishing [49]. Moreover, it is reasonable to neglect parity coupling, since the parities of SOTCS in Figs. 2c-2d is well-identified without mixing. Therefore, we assemble the two copies of parity-dependent Hamiltonian as $\hat{H}_{NN} = \hat{H}_{KM} \otimes \sigma_z$, where the Pauli matrix $\sigma_z$ represents the parity DoF. Applying the same approach on a finite-sized triangular lattice with $\kappa_{h1}/\kappa_{h2} < 1/2$, the calculated even/odd-parity SOTCS are degenerate at zero frequency, not consistent with the parity splitting in Fig. 2b. It implies that modelling the plasmonic lattice is not trivial as two copies of $\hat{H}_{KM}$, and indicates mechanisms beyond the NN coupling of hexapole mode.

We further reveal the underlying mechanism with a straight plasmonic trimer, which is a typical component in the Kagome lattice. The first-principal results in Fig. 2f show that each parity exhibits three nondegenerate supermodes around zero frequency. In these supermodes, hexapoles, which is *relevant* to our interest, dominate most cavities, thus meaning the dominating NN coupling ($\kappa_h$) of hexapole modes. Strikingly, in the middle supermodes of both parities, the middle resonator exhibits vanishing hexapoles but relatively weak quadrupoles (octupoles), which is *irrelevant* to our interest. It indicates the existence of the interorbital couplings (IOC) between hexapole and quadrupole. To investigate the effect of such IOC, a more accurate but complicated tight-binding model is required on a larger space including all orbitals. In the weak interacting scenario, it is feasible to transform a complicate model of higher dimensions to a reduced model by exploring Schrieffer-Wolff transformation, which has been widely used in condensed-matter [50-53]. Such Schrieffer-Wolff transformation has also been transferred to elastic systems [54] to extract the reduced model from simulation results. By transformation, the IOC is equivalent to effective parameters in the reduced model on the basis $[A_L \quad A_M \quad A_R]^T$ (where L, M, R denote hexapole modes on the left, middle and right cavities, respectively). The reduced model is generally presented as follows, and its NN coupling and effective effects of IOC are isolated for understanding:

$$\hat{H}_{\text{trimer}} = \underbrace{\begin{pmatrix} 0 & \kappa_h & 0 \\ \kappa_h & 0 & \kappa_h \\ 0 & \kappa_h & 0 \end{pmatrix}}_{\hat{H}_{\text{trimer}}^{NN}} + \underbrace{\begin{pmatrix} \Delta\omega & 0 & \kappa_3 \\ 0 & 2\Delta\omega & 0 \\ \kappa_3 & 0 & \Delta\omega \end{pmatrix}}_{\hat{H}_{\text{trimer}}^{IOC}}, \qquad (1)$$

where $\Delta\omega$ and $\kappa_3$ represent the IOC-induced effective on-site detuning (EOD) and long-range coupling (LRC), respectively. The EOD on the middle-cavity mode is twice of both the left- and right-cavity modes, since the middle-cavity mode couples with both the left and right cavity modes [49]. Such EOD originating from the inherent IOC, are different from the on-site detuning by changing metaatoms [41]. Regarding the effective LRC $\kappa_3$, the intuitive picture is given in Fig. 2g. A coupling channel emerges as (strong) hexapole - (weak) quadrupole - (strong) hexapole. The weak quadrupole in the middle resonator behaves as an effective bond, and couples the left and right hexapole modes. Noteworthy, the effective LRCs originating from IOC, are essentially different from the direct LRC induced by low[36].

We show the physical manifestations of IOC in the trimer by theoretically switching off/on $\Delta\omega$ and $\kappa_3$. In the ideal scenario with vanishing IOC, i.e., $\Delta\omega = \kappa_3 = 0$, the $\hat{H}_{\text{trimer}}$ gives three eigenmodes $\psi_- = \frac{1}{2}\begin{bmatrix} 1 & -\sqrt{2} & 1 \end{bmatrix}^T$, $\psi_2 = \frac{\sqrt{2}}{2}\begin{bmatrix} 1 & 0 & -1 \end{bmatrix}^T$, and $\psi_+ = \frac{1}{2}\begin{bmatrix} 1 & \sqrt{2} & 1 \end{bmatrix}^T$, at eigenfrequencies $\omega_- = -\sqrt{2}\kappa_h$, $\omega_2 = 0$, $\omega_+ = \sqrt{2}\kappa_h$, respectively. For subsequent comparisons, two quantities are crucial, i.e., the averaged eigenfrequencies $\bar{\omega} = (\omega_- + \omega_+)/2$, and the middle eigenfrequency $\omega_2$. In such ideal case, both $\bar{\omega}$ and $\omega_2$ are pinned at the zero frequency. We further switch on the IOC, i.e., $\Delta\omega \neq 0$, and $\kappa_3 \neq 0$. Due to the perturbative nature of the IOC, the eigenmodes are very close to those in the ideal case. In spectra, the IOC manifests as frequency shifts of both quantities ($\bar{\omega}$ and $\omega_2$), i.e., $\bar{\omega} = (3\Delta\omega + \kappa_3)/2$, and $\omega_2 = \Delta\omega - \kappa_3$. By simple mathematic manipulations, we obtain

$$\Delta\omega = \frac{1}{4}(2\bar{\omega} + \omega_2), \quad \kappa_3 = \frac{1}{4}(2\bar{\omega} - 3\omega_2), \quad \kappa_h \approx \frac{\sqrt{2}}{4}(\omega_+ - \omega_-). \qquad (2)$$

The first-principle calculated results in Fig. 2f show that $\bar{\omega}_{even} \approx -\bar{\omega}_{odd}$ and $\omega_{2,even} \approx -\omega_{2,odd}$. Substituting them into Eq. (2), the EOD and LRC are sign reversal for different parities, i.e. $\Delta\omega_{even} = -\Delta\omega_{odd}$, and $\kappa_{3,even} = -\kappa_{3,odd}$, thus indicating the sign reversal of the parity-dependent IOC. Such sign reversal of parity-dependent IOC is also consistent with the analysis of field overlap integral [49]. According to the first-principle calculated results in Fig. 2f, $\omega_+$ is larger (smaller) than $\omega_-$ for the odd (even) parity. We also have the sign reversal of $\kappa_h$ for different parities, i.e., $\kappa_{h,even} \cdot \kappa_{h,odd} < 0$.

The analysis above is further applied to the periodic Kagome lattice, whose Hamiltonian is expressed as

$$\hat{H}_1 = (\hat{H}_{KM} + \hat{H}_{IOC}) \otimes \sigma_z \quad (3)$$

where $\hat{H}_{KM}$ is the Hamiltonian of the conventional Kagome model, and the IOC term is expressed as

$$\hat{H}_{IOC} = -4\Delta\omega \cdot \mathbb{I} + 2\kappa_3 \cdot \text{diag}\left\{2\cos\left(\frac{d}{2}k_x\right)\cos\left(\frac{\sqrt{3}}{2}dk_y\right), \cos(dk_x) + \cos\left(\frac{d}{2}k_x - \frac{\sqrt{3}d}{2}k_y\right), \cos(dk_x) + \cos\left(\frac{d}{2}k_x + \frac{\sqrt{3}d}{2}k_y\right)\right\} \quad (4)$$

where $d$ is the lattice constant, $\mathbb{I}$ is the identity matrix. The EOD $\Delta\omega$ and LRC $\kappa_3$ is extracted from a plasmonic dimer and trimer, respectively [49]. This Hamiltonian gives six parity-dependent bands [49], whose bulk topologies are characterized by Wannier centres [26,49]. These non-zero Wannier centres give rise to parity-dependent SOTCS in parity-dependent bandgaps [49], unlike previous parity-less SOTCS.

We then explore a finite triangular Kagome model with $\kappa_1/\kappa_2 = 1/4$ and $\Delta\omega = -\kappa_3$ in Fig. 2h, whose corresponding periodic lattice has non-zero parity-dependent Wannier centres. It hosts two groups of parity-dependent SOTCS in the parity-dependent bandgaps as shown in Fig. 2i. The effect of parity-dependent IOCs on the SOTCS is revealed by varying $\kappa_3$. Two groups of parity-dependent SOTCS are degenerate when $\kappa_3 = 0$, while they split in the spectrum during increasing $\kappa_3$. Since the plasmonic lattice in Fig. 2a corresponds to the model with $\kappa_3 = 0.1\kappa_1$ in Fig. 2i, the theoretically obtained parity splitting is consistent with the results in Figs. 2b-2d.

Experimentally, we implement the Kagome lattice on a designer-plasmonic metasurface as shown in Fig. 3a. The resonator is a textured metal disk [49] with ultrathin thickness, which has been extensively demonstrated as a low-frequency analogy to the plasmonic cavity in nanooptics [48-49,55-58]. The hexapole is resonant at $\omega_h = 6.069$ GHz in individual metaatom. Such resonator exhibits out-of-plane radiations leakage ($\gamma_h = 0.01225$ GHz) not existing in the 2D plasmonic cavity. The leakage is uniform across the lattice and does not change bulk topologies. The distances between resonators are set as $d_1 = 29$ mm, $d_2 = 26$ mm, respectively. The corresponding coupling coefficients are extracted as $\kappa_{1;\text{even,odd}} = \mp 0.060 \text{GHz}$ $\kappa_{2;\text{even,odd}} = \mp 0.130 \text{GHz}$ [49]. We further show the IOC in a designer-plasmonic trimer with equal distances ($d = 27.5$ mm), by measuring its near-field transmission spectra. In Fig. 3b, the measured spectra exhibit parity-dependent frequency shifts of $\bar{\omega}$ and $\omega_2$ with respect to zero frequency, thus verifying the IOC in the designer-plasmonic trimer. According to the previous parameter-extraction approach, the parity-dependent IOCs are equivalently quantified with LRC $\kappa_{3;\text{even,odd}} = \mp 0.006$ GHz and EOD $\Delta\omega_{\text{even,odd}} = \pm 0.009$ GHz. Fig. 3c shows the middle supermodes of the trimer, where the middle resonator exhibits vanishing hexapoles but weak patterns of superpositions of quadrupole and octupole. It further confirms the existence of parity-dependent IOCs. On top of these extracted parameters, the eigenspectrum of the finite triangular-shaped sample is obtained (left panel of Fig. 3d). Although the even-parity SOTCS overlap with the odd-parity edge states, they still reside in the even-parity bandgaps. So do the odd-parity SOTCS.

We demonstrate the parity splitting of SOTCS using near-field spectral and imaging technologies. The right panel of Fig. 3d shows detected local densities of states (LDOS) at points B and A in Fig. 3a, where are the field maximum of odd(even) mode. The measured peaks are at 6.025 GHz (blue) and 6.134 GHz (red), respectively, corresponding well to the calculated eigenfrequencies of SOTCS. Near-field imaging results further confirm that the SOTCS at different resonance frequencies belong to specific parities. In Fig. 3e, the field pattern at 6.025(6.134) GHz corresponds to the odd(even)-parity SOTCS. Some edge lattices exhibit bright fields, but they belong to the even(odd) parity. This is consistent with the theoretical results in Fig 3d, where odd(even)-parity SOTCS overlap with the even(odd)-parity edge states in spectra. The robustness results of SOTCS are given in [49]. The experimental results for edge and bulk states are also given in [49].

The link between far-field polarizations and the near-field parities of SOTCS is further experimentally observed due to the existence of out-of-plane radiations in each lattice. The plasmonic HOTI is normally illuminated with $E_x$ and $E_y$ polarized waves, respectively. A vertical near-field probe is exploited to detect $E_z$ components at point A(B) of the corner resonator in Fig. 4a. This setup guarantees that the captured signals are clear from the SOTCS without the mixing from the incident waves. The vanished spectrum without metasurface confirms the setup. In Fig. 4b, the red(blue) spectrum detected at A(B) shows resonance peaks at 6.1(6.0) GHz and is consistent with both theoretical and near-field experimental results in Fig. 3d. Therefore, the red(blue) peaks in Fig. 4b imply that the $E_x$ ($E_y$) polarizations are projected to the even(odd) parity of SOTCS. The near-field patterns on the corner resonator in Figs. 4c(d) further verify the polarization-orthogonal SOTCS. The correspondences between the far-field polarizations and near-field parities are elucidated with the dipole decompositions shown in Fig. 4e and 4f. In an individual cavity, the hexapole can be decomposed into three dipoles with the same strength. Therefore, the total dipolar momentum vanishes. However, in the Kagome lattice, the hexapole on the corner resonator is perturbed by the coupling from its neighbours, inducing the nonequal strength of the three composite dipoles. As shown in Fig. 4e(f), the vertical (horizontal) dipole is stronger than the other two oblique ones (more analysis in [49]), thus leading to a net residual vertical (horizontal) dipole. Such residual dipoles are projected to far-field polarizations, thus giving rise to the polarization-orthogonal nondegenerate SOTCS, which are consistent with both theoretical and near-field experimental results.

In conclusion, we propose an approach to manipulate SOTCS by combining the orbital parity DoF and the interorbital coupling. By exploring designer-plasmonic metasurface, we experimentally demonstrated polarization-orthogonal nondegenerate SOTCS. In contrast to pseudospin DoF relying on $C_6$ or $C_3$ symmetries [43-45], the parity of hexapole is an intrinsic DoF that does not depend on any crystalline symmetries. Hence, intriguing topological physics could be anticipated by incorporating parities into other lattices. Since polarization-orthogonal nondegenerate cavity modes show significant applications in optimal single photon emitters [18], our finding could further endow topological protections to that application. By interfacing with gain media, our results may also promise applications in polarization-dependent topological lasing [46,47] and high-performance quantum emitters [59].

The work at Zhejiang University was sponsored by the Key Research and Development Program of the Ministry of Science and Technology under Grants No. 2022YFA1404902, 2022YFA1404704, and 2022YFA1405200, and the National Natural Science Foundation of China (NNSFC) under Grants No. 62171406, No.11961141010, No.61975176, No.62222115, No.62201500, No.U21A6006, the ZJNSF under Grant No. Z20F010018, the Fundamental Research Funds for the Central Universities No. 2020XZZX002-15, the Key Research and Development Program of Zhejiang Province under Grant No.2022C01036, and the Fundamental Research Funds for the Central Universities. This work at Jilin University was sponsored by NNSFC under Grants No. 62175083, No. 61935015, No. 61825502, the Natural Science Foundation of Jilin Province No. 20230101359JC.

**Figures**

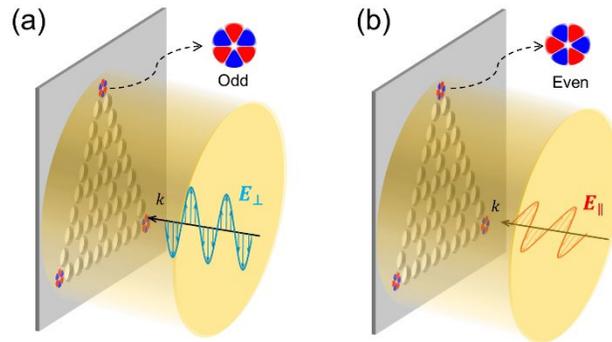

**Fig. 1. The schematic of polarization-orthogonal nondegenerate SOTCS on a plasmonic Kagome metasurface.** The blue/red waves with vertical/horizontal polarizations ($E_\perp$ / $E_\parallel$) correspond to the lower/higher frequency, respectively. The parities of SOTCS are shown in the insets.

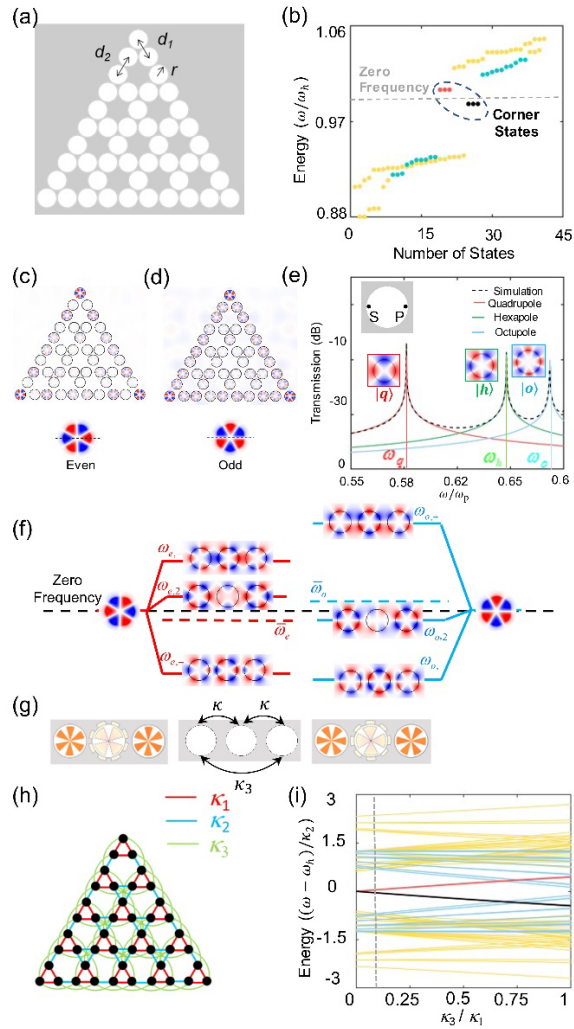

**Fig. 2. Parity-splitting of plasmonic SOTCS. a.** A two-dimensional plasmonic HOTI. $r$, $d_1$ and $d_2$ represent cavities radius and distances, respectively. **b.** Eigenfrequencies of the HOTI. The red(black) dots represent SOTCS with even(odd) parity. The grey dashed line marks the zero-frequency. **c-d.** Field patterns of SOTCS. **e.** The black dashed line denotes near-field transmission spectrum of a single cavity. The red, green, and blue solid lines are Lorentzian fittings. The "S(P)" mark the position of source(probe). **f.** The parity-dependent trimer supermodes, and their corresponding eigenfrequencies. The black dashed line represents the zero frequency. The red(blue) dashed lines represent the average frequency $\bar{\omega} = (\omega_- + \omega_+)/2$ of even(odd) modes. **g.** The schematic of the IOC-induced effective LRC.

**h.** The tight-binding model of the HOTI in (**a**). The $\kappa_1$, $\kappa_2$, and $\kappa_3$ are the intracell, intercell couplings, and effective LRC, respectively. **i.** The calculated eigenspectrum of the model in (**h**). The red(black) line denotes the eigenfrequencies of even(odd) SOTCS. The dashed line corresponds to the HOTI in (**b**).

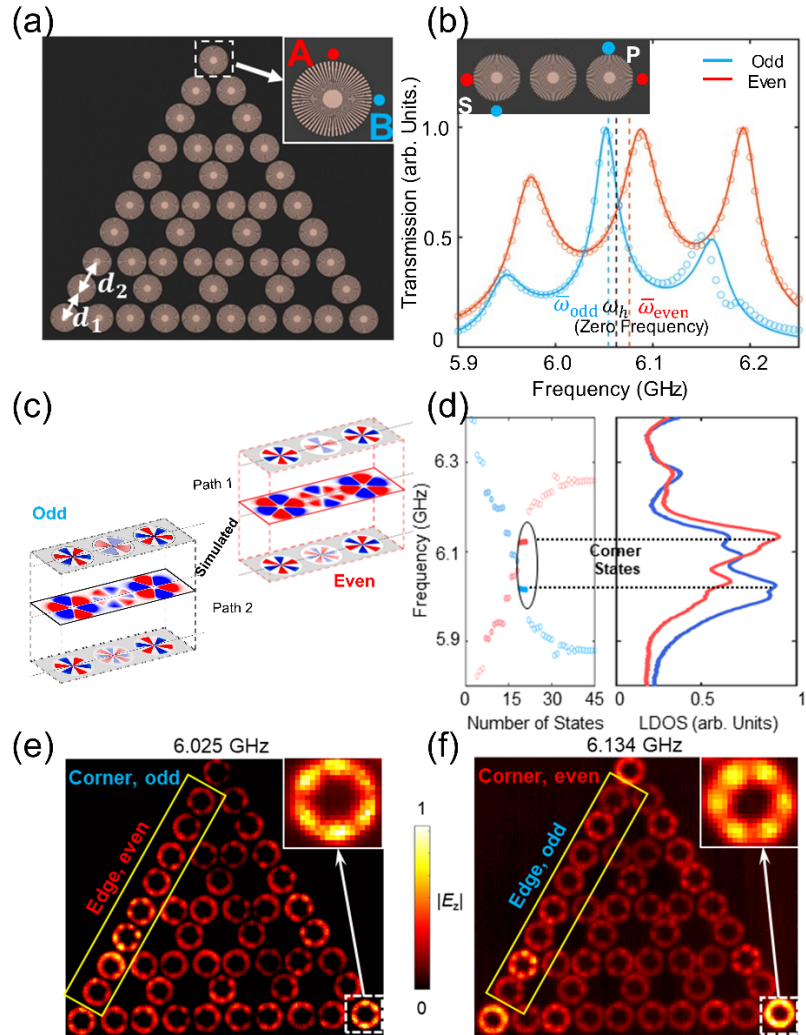

**Fig. 3. Demonstrating parity splitting of SOTCS on a designer-plasmonic metasurface.**
**a.** The photograph for the sample, with $d_1$ = 29 mm, and $d_2$= 26 mm. We probe even(odd) SOTCS at "A(B)". **b.** The transmission of a meta-trimer shown in the inset. "S(P)" denotes the source(probe). **c.** The field patterns at $\omega_{2;\text{odd}}$ ($\omega_{2;\text{even}}$). The patterns on the middle metaatom are the superpositions of quadrupole, and octupole. **d.** The calculated eigenfrequencies of the metasurface, and measured LDOS at A(B) in (**a**). The red(blue) corresponds to even(odd). **e-f.** The experimentally captured near-field patterns of SOTCS.

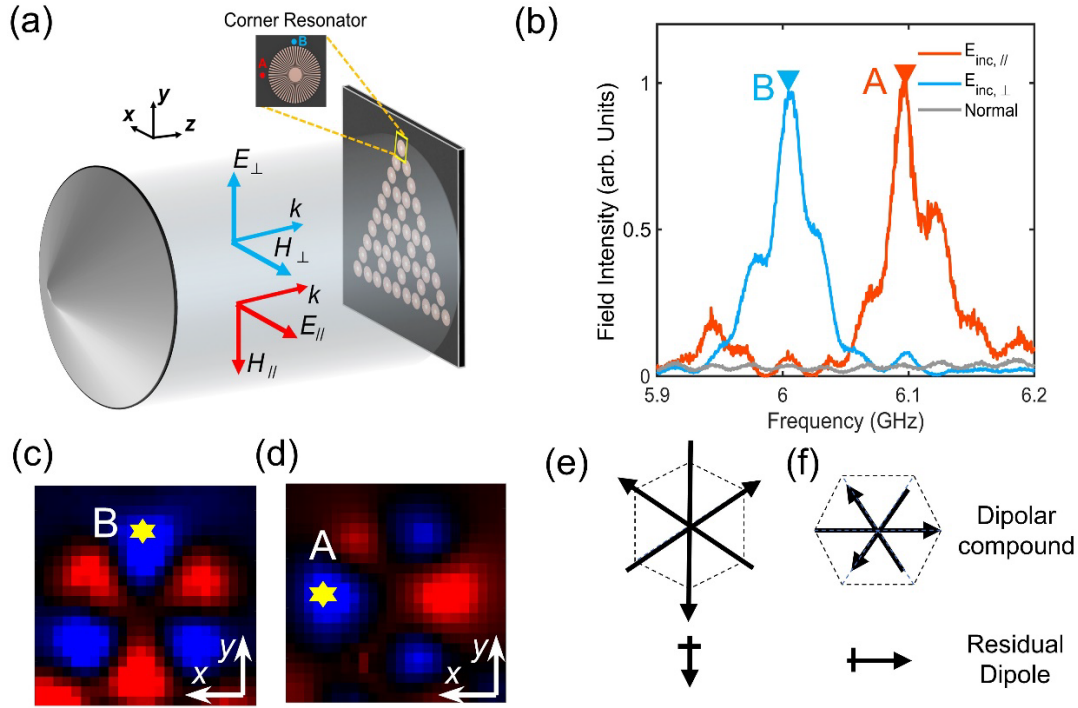

**Fig. 4. Observing polarization-orthogonal nondegenerate SOTCS. a.** The schematic of far-field experiment setup. The sample is normally illuminated with linearly-polarized beams. **b.** The measured spectra at A(B) in (**a**) corresponds to x(y)-polarized incidence. The grey line represents the spectrum without sample. **c-d.** Measured patterns on the top corner resonator at peak frequencies in (**b**). **e-f.** The polarization-parity correspondence. Dipolar compound of the hexapole on the corner resonator corresponding to the odd (**e**) and even (**f**) SOTCS.